%% file: CaLossesPaper.tex
\documentclass[fleqn,usenatbib]{mnras}

\usepackage{newtxtext,newtxmath}

\usepackage[T1]{fontenc}

\DeclareRobustCommand{\VAN}[3]{#2}
\let\VANthebibliography\thebibliography
\def\thebibliography{\DeclareRobustCommand{\VAN}[3]{##3}\VANthebibliography}

\usepackage{siunitx}

\usepackage{graphicx}	
\graphicspath{{./}}
\usepackage{amsmath}	

\newcommand*{\Caii}{Ca\,\textsc{ii}}
\newcommand*{\Caiii}{Ca\,\textsc{iii}}
\newcommand*{\CaLine}{Ca\,\textsc{ii} \SI{854.2}{\nano\metre}}

\newcommand*{\Lw}{\textit{Lightweaver}}

\DeclareSIUnit\erg{erg}

\title[Ca\,\textsc{ii} Photoionisation in Flare Simulations]{On the Importance of \Caii{} Photoionisation by the Hydrogen Lyman Transitions in Solar Flare Models}

\author[C. M. J. Osborne et al.]{
C. M. J. Osborne,$^{1}$\thanks{E-mail: c.osborne.1@research.gla.ac.uk}
P. Heinzel,$^{2}$
J. Kašparová,$^{2}$
and L. Fletcher$^{1,3}$
\\
$^{1}$SUPA School of Physics and Astronomy, University of Glasgow, Glasgow, G12 8QQ, UK\\
$^{2}$Astronomical Institute of the CAS, CZ-25165, Ondřejov, Czech Republic\\
$^{3}$Rosseland Centre for Solar Physics, University of Oslo, P.O. Box 1029 Blindern, NO-0135, Oslo
}

\date{Accepted XXX. Received YYY; in original form ZZZ}

\pubyear{2021}

\begin{document}
\label{firstpage}
\pagerange{\pageref{firstpage}--\pageref{lastpage}}
\maketitle

\begin{abstract}
\input{Header}
\end{abstract}




\input{Body}

\section*{Data Availability}

The data underlying this article is available in Zenodo at \url{https://dx.doi.org/10.5281/zenodo.4727772}. The \Lw{} framework (v0.5.0) is available in Zenodo at \url{https://dx.doi.org/10.5281/zenodo.4549258}, and the scripts associated with the model presented here is available in Zenodo at \url{https://doi.org/10.5281/zenodo.4757502} (v1.1.0).



\bibliographystyle{mnras}
\bibliography{Refs} 








\bsp	
\label{lastpage}
\end{document}

%% file: Header.tex
The forward fitting of solar flare observations with radiation-hydrodynamic simulations is a common technique for learning about energy deposition and atmospheric evolution during these explosive events. A frequent spectral line choice for this process is \CaLine{} due to its formation in the chromosphere and substantial variability. It is important to ensure that this line is accurately modeled to obtain the correct interpretation of observations. Here we investigate the importance of photoionisation of \Caii{} to \Caiii{} by the hydrogen Lyman transitions; whilst the Lyman continuum is typically considered in this context in simulations, the associated bound-bound transitions are not. This investigation uses two RADYN flare simulations and reprocesses the radiative transfer using the \Lw{} framework which accounts for the overlapping of all active transitions. The \CaLine{} line profiles are found to vary significantly due to photoionisation by the Lyman lines, showing notably different shapes and even reversed asymmetries.
Finally, we investigate to what extent these effects modify the energy balance of the simulation and the implications on future radiation-hydrodynamic simulations. There is found to be a 10-15\% change in detailed optically thick radiative losses from considering these photoionisation effects on the calcium lines in the two simulations presented, demonstrating the importance of considering these effects in a self-consistent way.

%% file: Body.tex
\section{Introduction}

Radiation-hydrodynamic (RHD) simulations are a common approach to modelling solar flares against which observations can be compared, and theoretical predictions investigated.
The most commonly used of these are RADYN
\citep{Carlsson1992a,Carlsson1997,Abbett1999,Allred2005,Allred2015}, FLARIX
\citep{Varady2010,Heinzel2015a}, and HYDRAD \citep{Bradshaw2003,Bradshaw2013,Reep2019}.
These models are field-aligned, viewing the plasma as a quasi-one-dimensional
field-aligned tube of fluid, coupled with a plane-parallel description of
non-LTE (NLTE) radiative transfer. RADYN and FLARIX both perform full
non-equilibrium (i.e. time-dependent) computation of hydrogen and calcium
level populations, ionisation states, and spectral synthesis (RADYN also
treats helium in this way, and both have been adapted to treat magnesium).
Recently, RADYN and FLARIX simulations have been compared, and found to have
good agreement \citep{Kasparova2019}.

In this work we reprocess the radiative transfer aspect of RADYN simulations
with time-dependence to investigate the impact of \Caii{} to \Caiii{}
photoionisation by the hydrogen Lyman lines, an effect that is not considered
in RADYN's current radiative treatment, in addition to the Lyman continuum which is considered in both models.
If these lines have a significant photoionising effect then the distribution of calcium populations between ionisation states will be affected, along with the observed line profiles and even the radiative losses from the atmosphere.
This effect was first modelled and discussed by \citet{Ishizawa1971} and is an important component of prominence modelling \citep[e.g.][]{Gouttebroze2002}.
The \CaLine{} spectral line of the infrared triplet is chosen for this
investigation as it is a strongly variable chromospheric line commonly
treated in RHD codes due to the importance of calcium transitions in
determining atmospheric energy balance, making it a prime candidate for
comparison and forward-fitting of observations (e.g. \citet{Kuridze2015,
Kuridze2018, Kerr2016, RubioDaCosta2016, Bjorgen2019}).
\CaLine{} has strong diagnostic potential and is observed in high spatial,
spectral, and temporal resolution using instruments such as CRISP on the
Swedish Solar Telescope (SST) \citep{Scharmer2008}, and the upcoming Visible
Tunable Filter (VTF) on the Daniel K Inouye Solar Telescope (DKIST)
\citep{Rimmele2020}.
It can therefore be used to probe thermodynamic properties in the
chromosphere, primarily centred on the core-forming region at an average
optical depth $\log \tau_{\SI{500}{\nano\metre}} \sim -5.3$ \citep{Centeno2018}.
This information can be augmented with observations of other spectral lines such
as H$\alpha$ to better understand thermodynamic gradients present in the
solar atmosphere and interpret energy deposition and transport during solar
flares.
Whilst flare observations typically consider \CaLine{} in an unpolarised
mode, the line is highly polarisable and thus can carry an additional
wealth of information regarding the chromospheric magnetic field
\citep{Centeno2018}.
To correctly interpret the large volume of high quality observations that
will be taken over the coming years, using forward-fitting
\citep[e.g.][]{Kuridze2015,RubioDaCosta2016}, traditional regression-based inversions
\citep[e.g.][]{Kuridze2018}, and more efficient machine-learning approaches
\citep{Osborne2019} it is essential that RHD models synthesise
\CaLine{} as accurately as possible.

The hydrogen Lyman lines are very strongly
enhanced in observations of flares, where a two order of magnitude
enhancement between quiet sun and flaring region was found by
\citet{RubioDaCosta2009} using the Transition Region and Coronal Explorer
(TRACE), and also in RHD simulations using RADYN which suggest similar or even
larger enhancements \citep{Brown2018,Hong2019}.
These highly enhanced lines lie in a wavelength range spanned by several
\Caii{} to \Caiii{} continua and therefore provide a mechanism for
\Caii{} photoionisation, possibly influencing the opacity throughout the chromosphere
and in turn the energy balance and emergent calcium spectral line profiles.

We will first present the methodology of treating the Lyman lines together
with the calcium transitions in these simulations, followed by the effects of
this photoionisation on the emergent line profiles. Finally, we will
investigate whether these effects change the \Caii{} radiative losses
sufficiently to modify the energy balance of the model.

\section{Methodology}

In this investigation we compare the line profiles and radiative losses from
\Caii{} spectral lines in RADYN simulations and those same simulations
reprocessed using the \Lw{} framework \citep{Osborne2021},
both with and without the photoionisation effect of the hydrogen Lyman lines.
First, a baseline simulation is produced using a slightly modified version of
the RADYN code in which the hydrodynamic variables and beam heating rates are
written to a file at every internal timestep. This time-dependent atmosphere is
then loaded into a program built on the \Lw{} framework and two simulations are
performed, one including the photoionising effects of the Lyman lines on \Caii{}
and the other excluding these effects (retaining photoionisation from the Lyman
continuum in these baseline models). We stress the \emph{framework} term attached
to \Lw{}, as it is not a code, but instead a modular set of components that can
be connected in different ways to construct a program specific to the
radiative transfer problem at hand, allowing for more rapid experimentation with
different techniques.  Similarly to RADYN, our simulations are
undertaken in a fully time-dependent manner whereby an initial solution for the
active atomic populations is computed in statistical equilibrium, the populations
are then advected, and the atmospheric properties are updated to the
thermodynamic atmosphere from the next RADYN timestep (including the calculation
of LTE populations and background opacities). Finally the populations are
advanced in time given the new atmospheric properties, updated radiation field,
and the atomic populations at the previous timestep.  These steps (other than
the initial statistical equilibrium) are taken for every timestep saved from the
RADYN simulation using RADYN's internal timestep.

The simulations presented here make use of RADYN in a similar configuration to that used for the F-CHROMA grid of simulations\footnote{Produced by the F-CHROMA project and available from
\url{https://star.pst.qub.ac.uk/wiki/doku.php/public/solarmodels/start.}}.
The starting atmosphere is based on the semi-empirical VAL3C model \citep{Vernazza1981}.
Instead of the Fokker-Planck formalism used in the F-CHROMA grid, we use the simpler analytic ``Emslie'' beam approach \citep{Emslie1978} with a spectral index $\delta=5$, low-energy cut-off of \SI{20}{\kilo\electronvolt}, and a constant energy deposition for \SI{10}{\s} of $1\times10^6$ or \SI{1e7}{\J\per\square\m\per\s} \footnote{In more commonly used cgs units for
RADYN simulations, these correspond to $1\times10^9$ and
\SI{1e10}{\erg\per\square\cm\per\s}.}.
Henceforth these two simulations with different energy depositions will be referred to as F9 and F10 respectively.
These parameters were chosen to serve as ``average'' RADYN simulations and are the same as those used by \citet{Kerr2019,Kerr2019a}.
They are also in agreement with the F-CHROMA grid, which uses spectral indices in the range of 3-8, low energy cut-offs in the range 10-\SI{25}{\kilo\electronvolt}, and total energy depositions in the range $3\times10^{7}$ to \SI{1e9}{\J\per\square\m} with triangular heating pulses lasting \SI{20}{\s}.

Observationally, our low-energy cut-off is supported by \citet{Sui2007}, who found a range of 10-\SI{50}{\kilo\electronvolt} in a sample of 33 early impulsive flares using the Ramaty High Energy Solar Spectroscopy Imager (RHESSI) assuming a cold collisional thick target model and accounting for X-ray albedo.
Our choice of spectral index is supported by \citet{Saint-Hilaire2008}, whose survey of 53 flares using RHESSI found a distribution of photon spectral index $\gamma$ in the range 2 to 5, peaking between 3 and 3.5.
The photon spectral index is related to the electron index by $\delta = \gamma+1$, thus our choice of spectral index lies well inside their distribution.

Our beam energy fluxes are on the lower end of those used in the F-CHROMA grid, in large part due to the constant energy deposition, which is much more demanding on the simulation due to the lack of beam ramp-up time.
\citet{Kerr2019} also present an F11 simulation (with otherwise identical parameters), but use a different version of RADYN with a different starting atmosphere, model atoms, and a Fokker-Planck model for the beam energy deposition.
We were unable to make a model with such high energy deposition converge, despite repeated adjustments to RADYN's parameters and several thousand hours of CPU time.
Additionally, both RADYN's and \Lw{}'s methods for including incoming optically thin coronal radiation were disabled due to discrepancies between their implementations.

The methods used in the \Lw{} framework are discussed in depth in its associated
technical report \citep{Osborne2021}. The framework is
numerically similar to the RH code \citep{Uitenbroek2001,Pereira2015}, using the
multi-level accelerated lambda iteration (MALI) method with full preconditioning
is used \citep{Rybicki1992}, along with a cubic Bézier spline formal solver
\citep{DelaCruzRodriguez2013}.  These techniques allow for multiple overlapping
lines and continua to be treated directly.
Partial redistribution effects on spectral lines can also be considered, however
this is not applied in this work, so as to maintain equivalency with RADYN in this
aspect of the treatment.
The populations are advanced in time following the same approaches
based on MALI preconditioning presented in \citet{Kasparova2003} and
\citet{Judge2017}. The advective term of the kinetic equilibrium equations is
also treated here, externally to the \Lw{} framework.

The kinetic equilibrium equations are solved by splitting the time-evolution
operator for the populations into radiative and advective components. The
radiative component is solved as previously discussed, over the entire
timestep, followed by the advective component. As RADYN's grid is also
adopted by our program, we also apply a modified version of its method of
advection. The advection equations for the populations use piecewise limited
reconstruction (van Leer style with second-order correction) and the problem
is then cast as an implicit system of ordinary differential equations solved
by Newton-Raphson iteration. The Jacobian needed for this is computed by
coloured finite-difference \citep{Curtis1974} on the residual to be
minimised. A Strang splitting \citep{Strang1968} approach coupling the
solutions of the advective and radiative terms to second order was also used,
but yielded no apparent differences in the result.


Our program independently computes the level populations and outgoing radiation
from hydrogen and calcium in the RADYN simulation.  A simulation is first run
with all hydrogen and calcium lines included, to consider the effects of
irradiation from the Lyman lines. A second simulation with a model hydrogen atom
excluding the resonance lines is then run, using fixed hydrogen level
populations loaded at each timestep from the data saved from RADYN (the
so-called ``detailed static'' mode in \Lw{});
the full time-dependent treatment is still applied to the calcium
populations.  In all of these simulations, for both RADYN and our model, the
Lyman continuum and its photoionisation effects on \Caii{} are
included. Non-thermal collisional processes in hydrogen are included using the
method of \citet{1993Fang}, using the beam heating rates saved from RADYN.

The hydrogen, calcium, and helium model atoms used in our model are the same as
those used in RADYN, and the other species used in LTE for the background are
taken from the RH distribution.  Therefore the hydrogen atom used has five bound
levels and and an overlying H\,\textsc{ii} continuum with ten lines, and the
\Caii{} atom also has five bound levels and an overlying \Caiii{} continuum with
five lines.  RADYN uses an approximate treatment for partial redistribution in
the Lyman lines by removing radiative broadening and reducing the van der Waals
broadening parameters in these lines. Whilst a more correct treatment will
somewhat change the intensity in these lines, and thus the \Caii{} populations,
we choose to replicate this treatment for consistency. The \Caii{} lines are
also affected by partial redistribution effects (primarily in the H \& K
resonance lines), but this is not considered in RADYN and should remain a small
effect in the high chromospheric densities that occur during flares.  In all
cases the same \SI{2}{\kilo\metre\per\second} microturbulent velocity is assumed
throughout the atmosphere.  The \Lw{} and RADYN models therefore differ only in
their prescriptions of background opacities (including the treatment of helium
in LTE in the \Lw{} model) and numerical techniques.

\begin{figure*}
    \centering
    \includegraphics[width=0.8\textwidth]{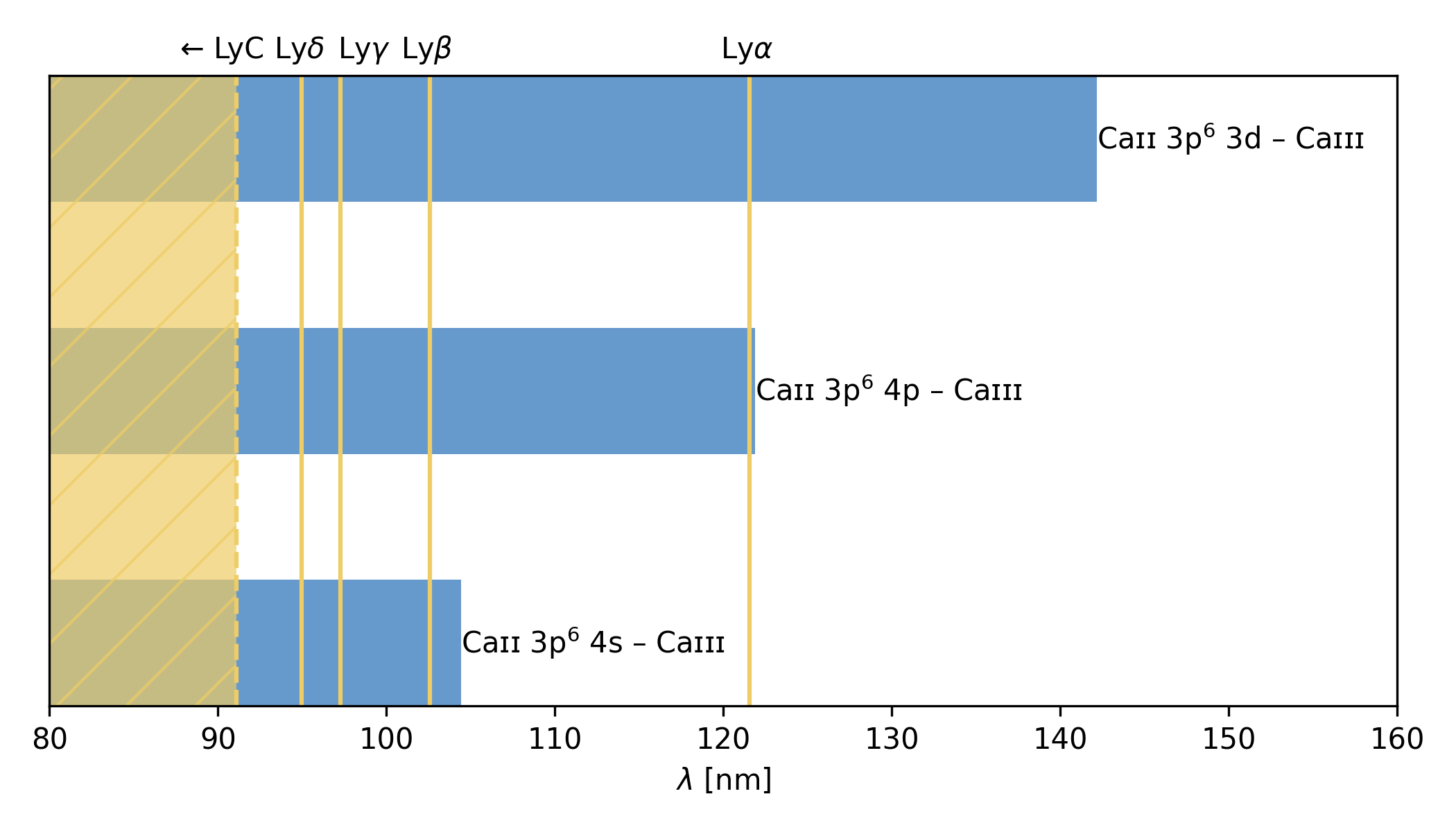}
    \caption{This diagram shows the overlap between the hydrogen Lyman lines and continuum with the \Caii{} continua present in our model. Both the \Caii{} 3p$^\mathrm{6}$ 4p and 3p$^\mathrm{6}$ 3d levels contain two sub-levels with close to identical continuum edges.}
    \label{Fig:OverlapDiagram}
\end{figure*}

Fig. \ref{Fig:OverlapDiagram} shows the overlap from the hydrogen Lyman lines
and continuum with the calcium continua present in our model.  Radiation from
Ly$\alpha$ photoionises \Caii{} from all levels present other than the ground
state, and the higher energy radiation from higher Lyman lines can photoionise
\Caii{} to \Caiii{} from all \Caii{} levels in this model. In reality, higher
Lyman lines, significantly Stark-broadened by the flaring atmosphere, will
create a quasi-continuum from a point between Ly$\delta$ and the Lyman continuum
\citep{DeFeiter1975}, further enhancing the photoionisation of \Caii{} from what
is presented here.

\section{Changes to line profiles}

\begin{figure*}
    \centering
    \includegraphics[width=0.7\textwidth]{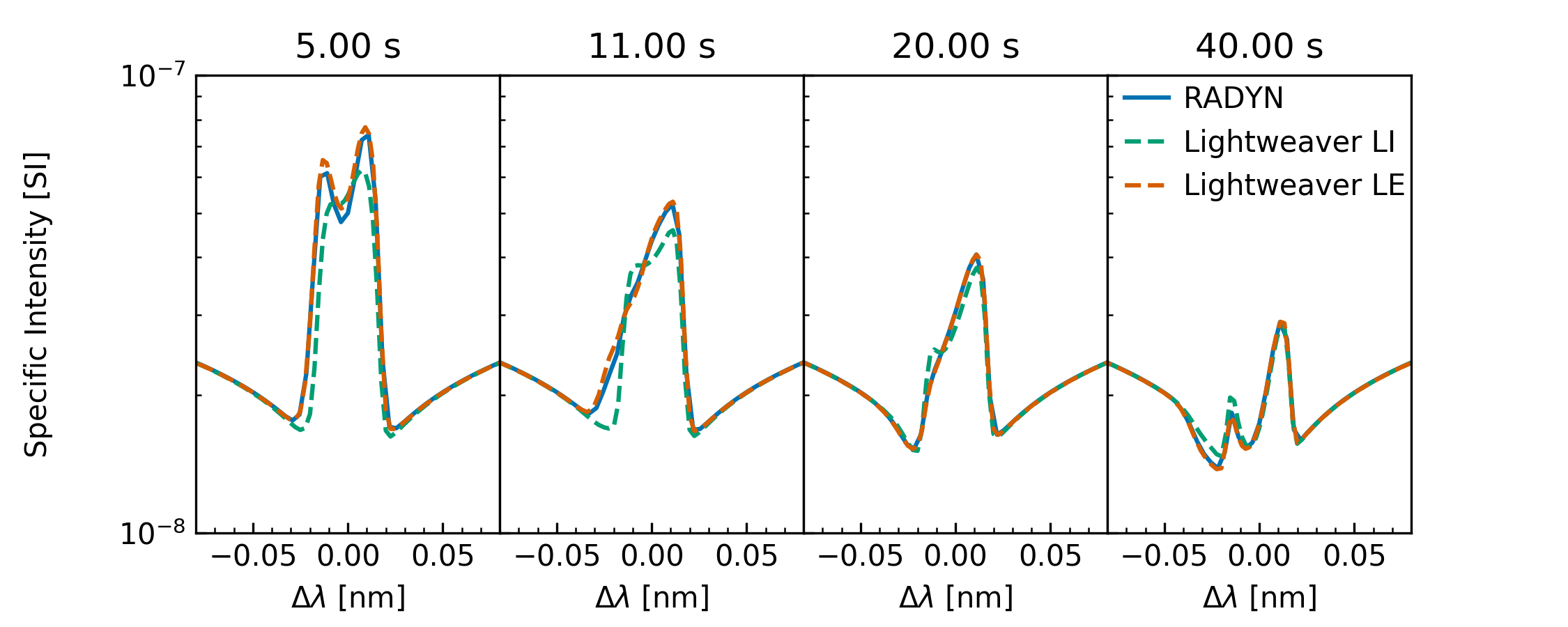}
    \caption{Comparison of \CaLine{} line profiles during the F9 simulation. Specific intensity is expressed in \si{\J\per\s\per\square\metre\per\Hz\per\steradian}. Here ``LI'' indicates the model where photoionisation of \Caii{} by the hydrogen Lyman lines is taken into account whereas ``LE'' excludes this effect.}
    \label{Fig:F9Profiles}
\end{figure*}%
\begin{figure*}
    \centering
    \includegraphics[width=0.7\textwidth]{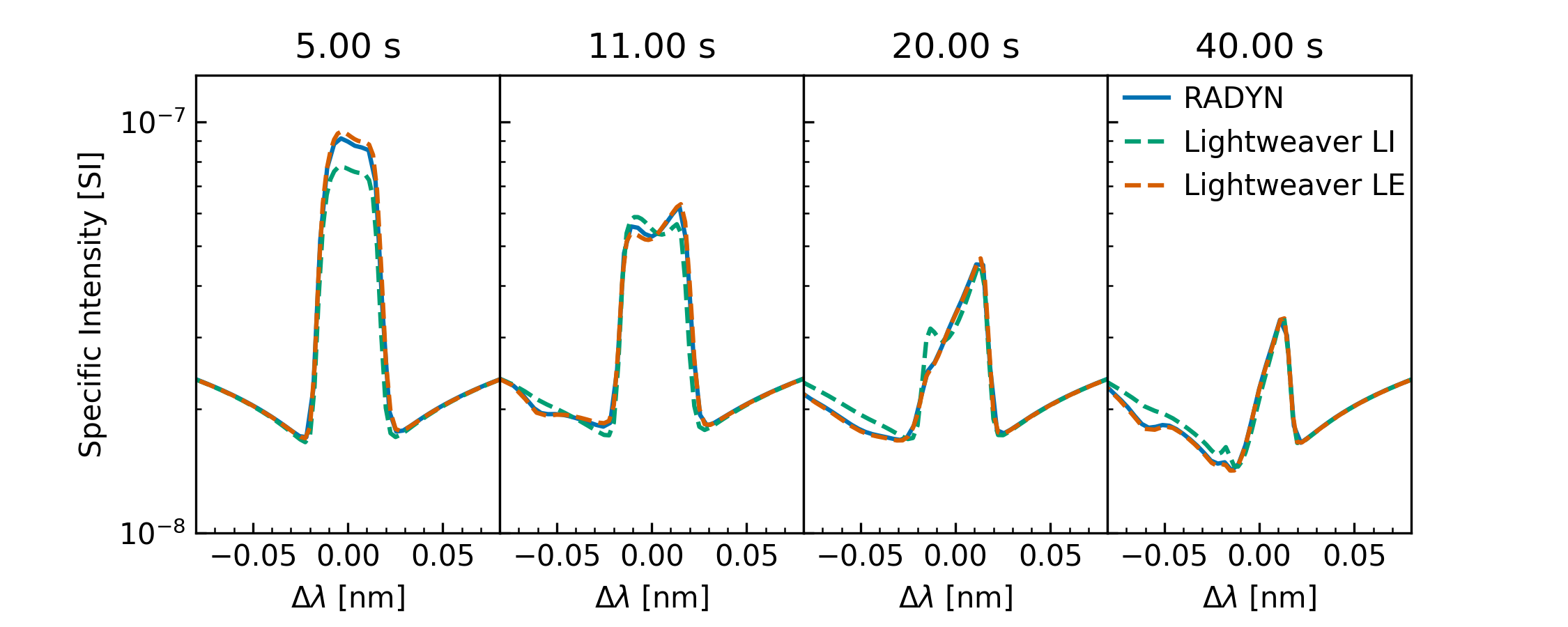}
    \caption{Comparison of \CaLine{} line profiles during the F10 simulation.}
    \label{Fig:F10Profiles}
\end{figure*}%

\begin{figure*}
    \centering
    \includegraphics[width=0.8\textwidth]{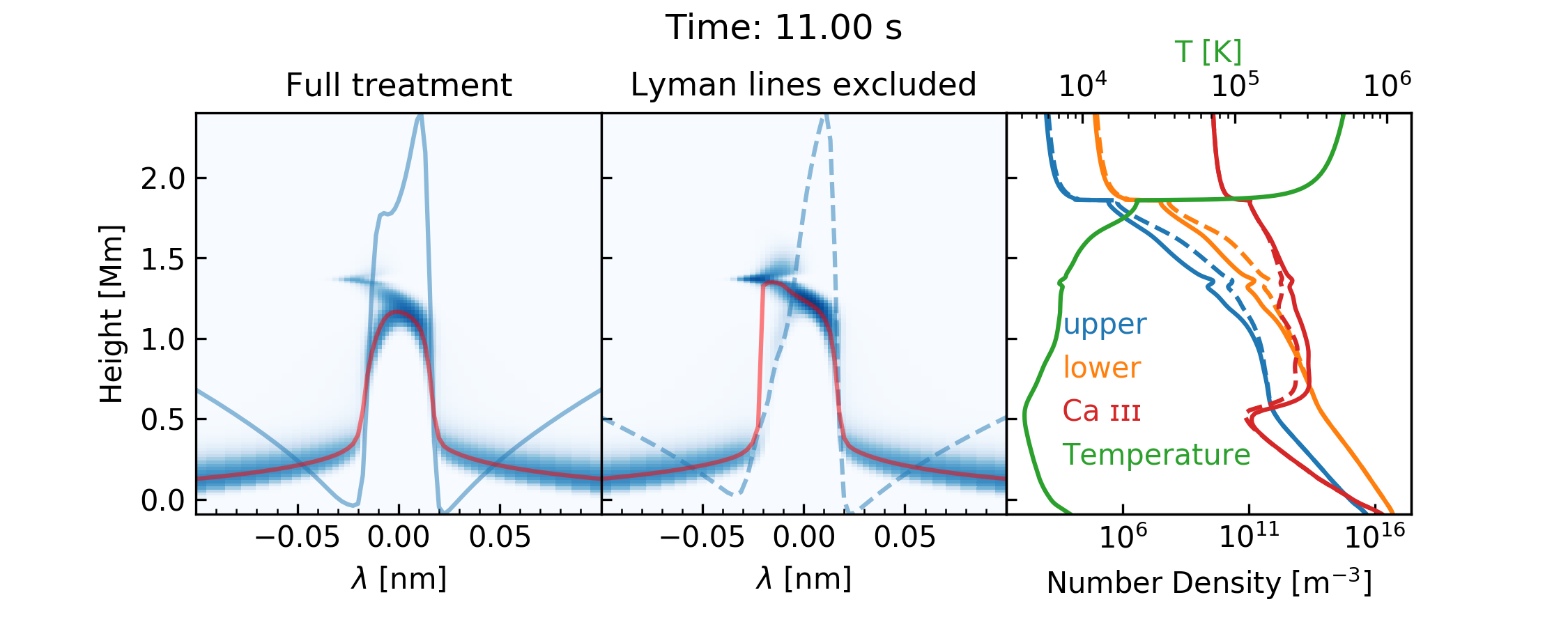}
    \caption{Contribution function and level populations for the two cases in the F9 simulation at $t=\SI{11}{\s}$. The first two panels show the contribution function, emergent line profile in blue, and $\tau_\nu = 1$ line in red. The third panel shows the temperature structure, and population density for upper and lower levels of the transition and \Caiii{}, where the solid lines represent the populations from the LI treatment and the dashed lines the LE treatment.}
    \label{Fig:F9Cfn11}
\end{figure*}
\begin{figure*}
    \centering
    \includegraphics[width=0.8\textwidth]{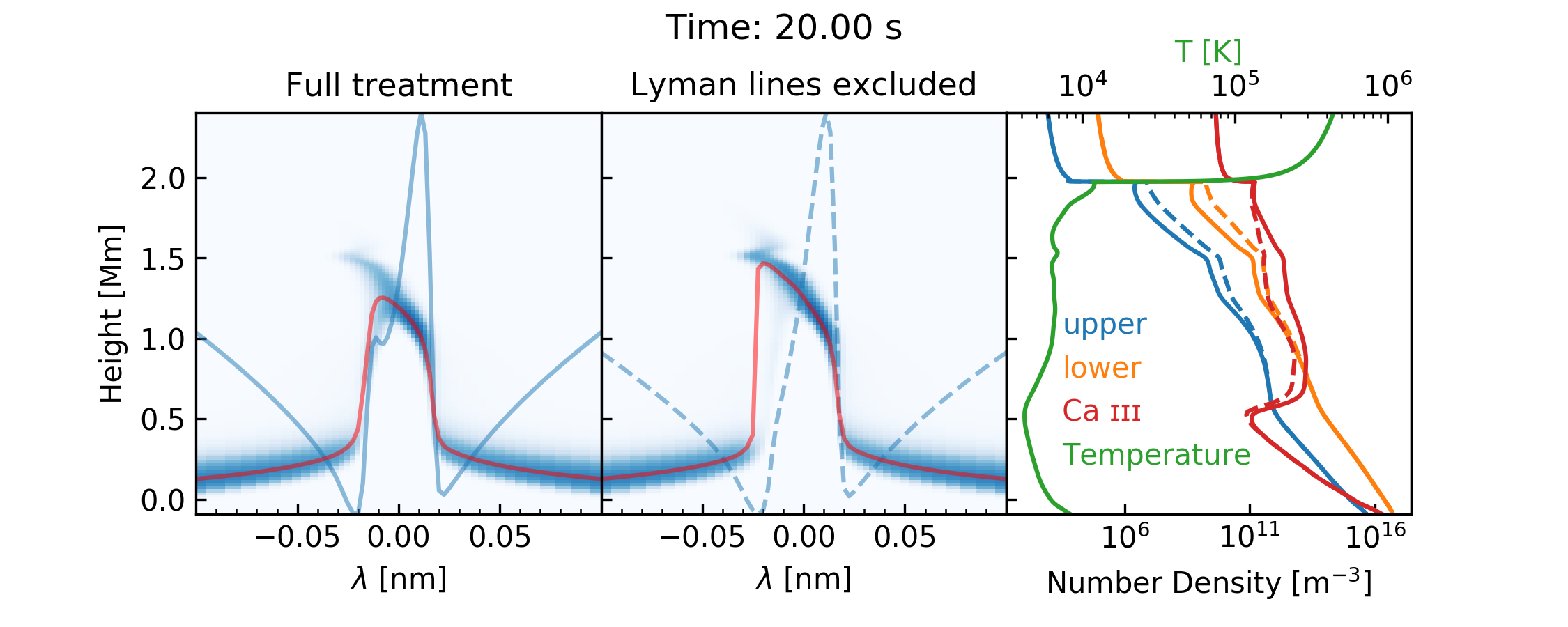}
    \caption{Contribution function and level populations for the two cases in the F9 simulation at $t=\SI{20}{\s}$. The panels represent the same information as Fig. \ref{Fig:F9Cfn11}.}
    \label{Fig:F9Cfn20}
\end{figure*}
\begin{figure*}
    \centering
    \includegraphics[width=0.8\textwidth]{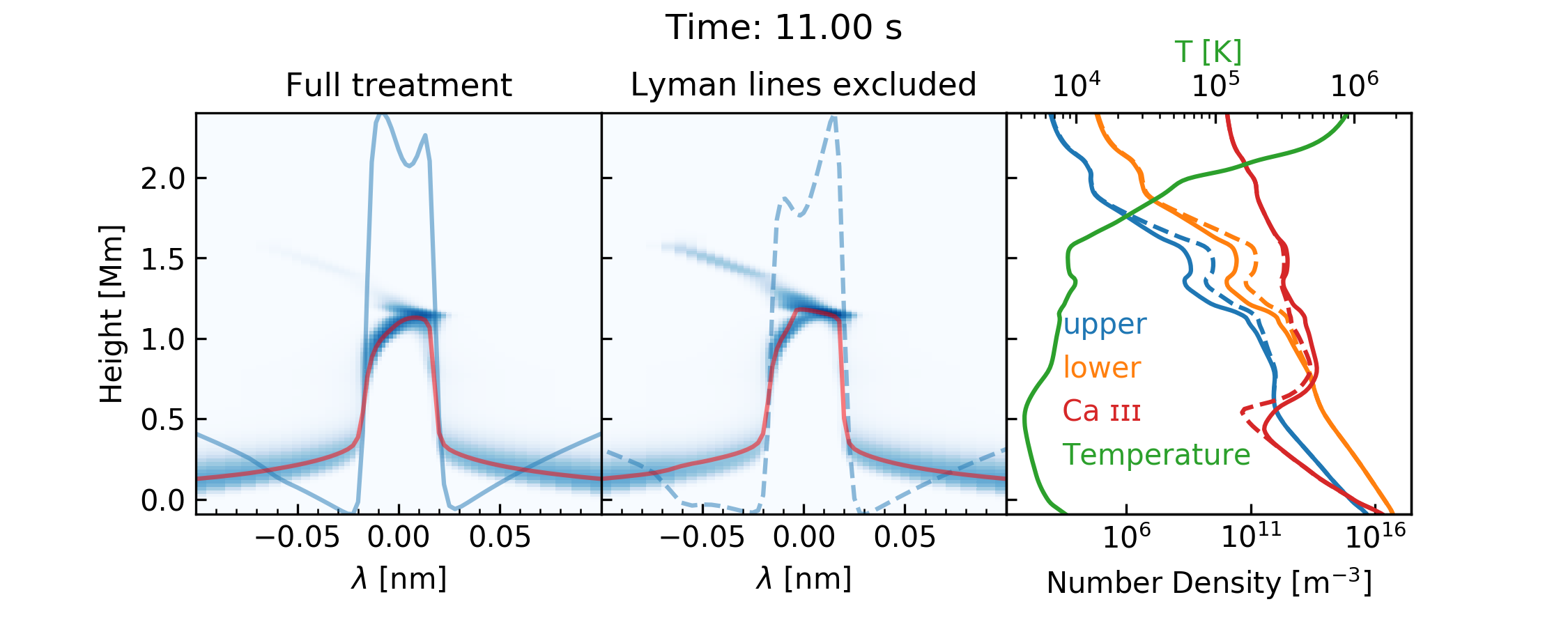}
    \caption{Contribution function and level populations for the two cases in the F10 simulation at $t=\SI{11}{\s}$. The panels represent the same information as Fig. \ref{Fig:F9Cfn11}.}
    \label{Fig:F10Cfn11}
\end{figure*}
\begin{figure*}
    \centering
    \includegraphics[width=0.8\textwidth]{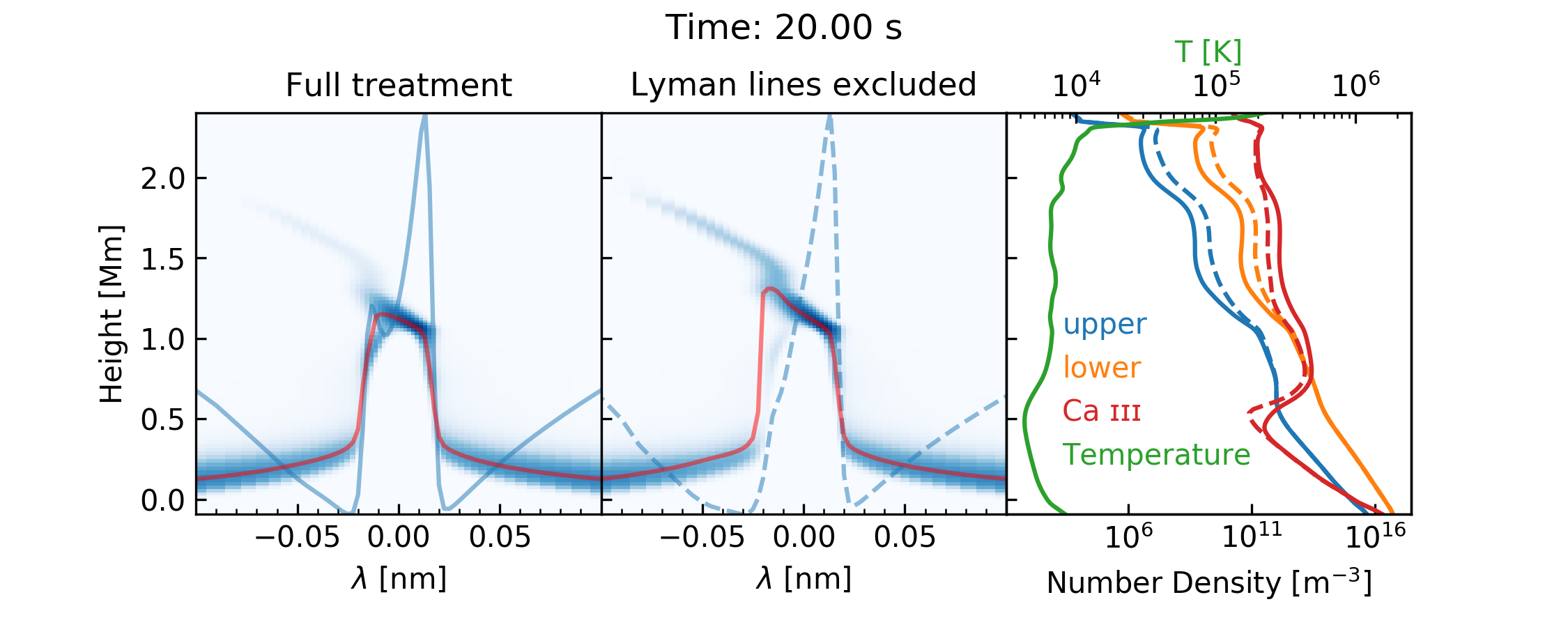}
    \caption{Contribution function and level populations for the two cases in the F10 simulation at $t=\SI{20}{\s}$. The panels represent the same information as Fig. \ref{Fig:F9Cfn11}.}
    \label{Fig:F10Cfn20}
\end{figure*}

In the following we describe the simulations where the effects of the
Lyman lines on \Caii{} are included as Lyman inclusive (LI), and the others
as Lyman exclusive (LE).


In Figs. \ref{Fig:F9Profiles} and \ref{Fig:F10Profiles} we show the \CaLine{}
line profile at $5$, $11$, $20$ and \SI{40}{\s} since the start of the
heating for both of the simulations considered, compared against the output
from the RADYN simulations. We see that the intensity from the LE model
disagrees with the RADYN output by only a few percent, and the shapes remain
extremely consistent, giving confidence in the \Lw{} framework. On the other hand, the line profiles produced by the LI
model differ remarkably from those produced by RADYN.

For the F9 simulation shown in Fig. \ref{Fig:F9Profiles}, the full LI treatment
produces a narrower, less intense, double-peaked line profile, whilst the LE
treatment produces a much more variable line profile, starting off sharply
double-peaked, and becoming singly-peaked in the decay phase (between $11$ and
\SI{20}{\s}).

For the F10 simulation shown in Fig. \ref{Fig:F10Profiles}, we note that
during flare heating the peak intensity is substantially reduced in the LI
treatment ($t=\SI{5}{\s}$). At $t=\SI{11}{\s}$ the asymmetry between the
two peaks is reversed in the LI and LE treatments, and at $t=\SI{20}{\s}$
there is a secondary peak in the violet side of the line profile in the LI case
that is not present in the LE treatment. At later times there are also dips to
lower intensity in the violet wing of the LE case, which evolve in shape over
this cooling phase.

Due to the different line asymmetries and presence of extra peaks, we
consider that the differences are most stark for the $t=11$ and
$t=\SI{20}{\s}$ timeslices in these simulations.  Looking at the level
populations for these in the right-hand panels of Figs.
\ref{Fig:F9Cfn11}-\ref{Fig:F10Cfn20} we see that the LE and LI models have
arrived at different populations for the upper and lower levels of \CaLine{}.
The contribution functions, shown in the left-hand panels of Figs.
\ref{Fig:F9Cfn11}-\ref{Fig:F10Cfn20}, also differ significantly, leading not
only to different line profiles, but to a different theoretical interpretation
of where the emergent line profile is formed. In the LE models there is a
contribution to the violet wing of the line present around \SI{1.5}{\mega\metre}
that is much less significant in the LI treatment. We suggest that this occurs
due to the lower population of levels involved in the formation of \CaLine{} at
this height, primarily due to the photoionisation of \Caii{} to \Caiii{} in this
region from Lyman lines formed in the transition region. This in turn reduces
the local opacity in an area that would otherwise, due to local plasma
temperature, be expected to be optically thick.  This interpretation is
supported by the change in the $\tau_\nu=1$ line present in the contribution
function figures. It is also interesting to note the large variation in \Caiii{}
populations in the F10 simulation around the temperature minimum region, much
deeper than where the cores of any of the \Caii{} lines form, likely due to
photoionisation from the upper chromosphere and transition region.

\section{Changes to radiative losses}

\begin{figure*}
    \centering
    \includegraphics[width=0.8\textwidth]{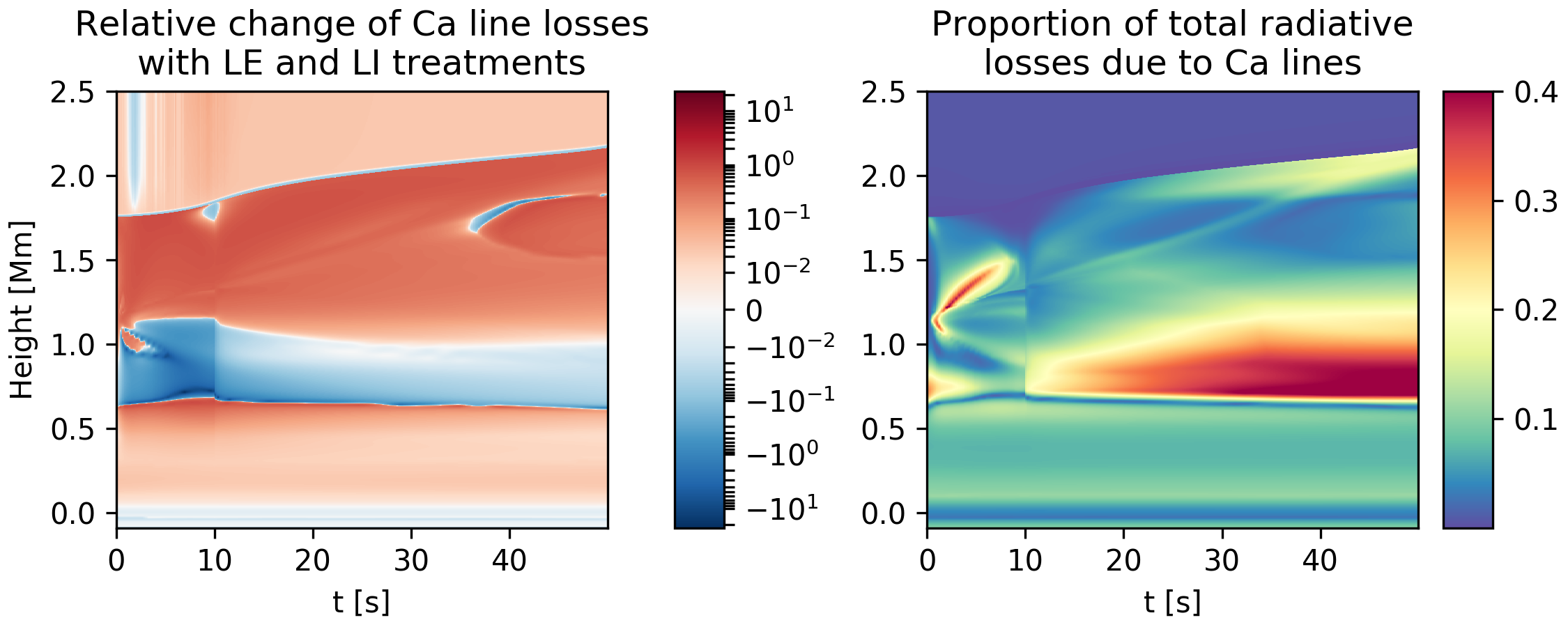}
    \caption{These panels show the time-evolution of the calcium losses in the F9 simulation. The left-hand panel shows the absolute relative change in losses due to calcium lines between the LI and LE models. The right-hand panel indicates the proportion of the total radiative losses due to calcium lines.}
    \label{Fig:Losses1e9}
\end{figure*}

\begin{figure*}
    \centering
    \includegraphics[width=0.8\textwidth]{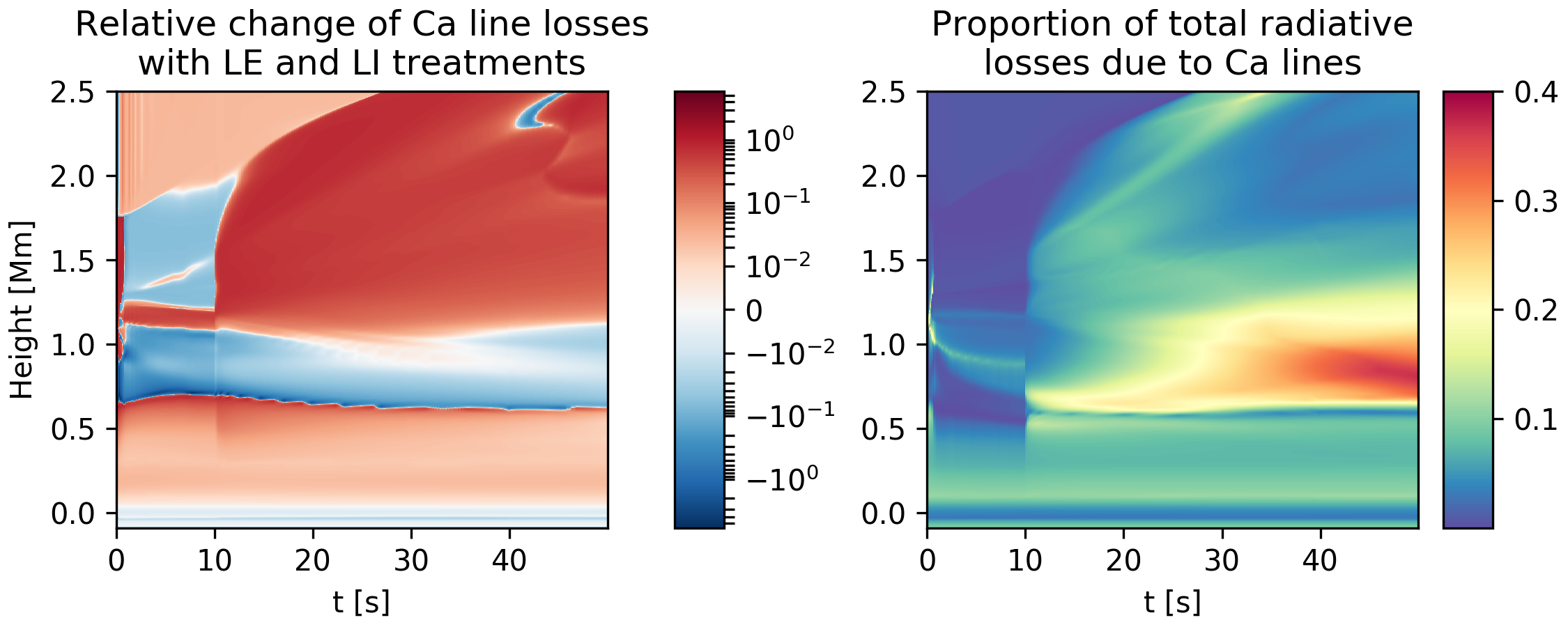}
    \caption{These panels show the time-evolution of the calcium losses in the F10 simulation. The left-hand panel shows the absolute relative change in losses due to calcium lines between the LI and LE models. The right-hand panel indicates the proportion of the total radiative losses due to calcium lines.}
    \label{Fig:Losses1e10}
\end{figure*}

Figs. \ref{Fig:Losses1e9} and \ref{Fig:Losses1e10} show the variation between the two calcium treatments in both simulations, as well as the proportion of the radiative losses that the calcium lines represent in the RADYN simulation. The fine band where there is a large difference between the two treatments is around the temperature minimum region of the original simulation, and should remain relatively unimportant for the formation of this line.
The left-hand panel is computed as
\begin{equation}
    \frac{\sum_{i\in \mathcal{C}} |\mathrm{loss}_{i,\, \mathrm{LE}}| - \sum_{i\in \mathcal{C}} |\mathrm{loss}_{i,\, \mathrm{LI}}|}{\sum_{i\in \mathcal{C}} |\mathrm{loss}_{i,\, \mathrm{LE}}|},
\end{equation}
where $\mathcal{C}$ is the set of calcium lines in our model, $\mathrm{loss}_i$ is the volumetric radiative loss (\si{\watt\per\metre\cubed}), and the right-hand panel is computed as
\begin{equation}
    \frac{\sum_{i\in \mathcal{C}} |\mathrm{loss}_{i,\, \mathrm{RADYN}}|}{\sum_j |\mathrm{loss}_{j,\, \mathrm{RADYN}}|},
\end{equation}
i.e. the fraction of the so-called detailed radiative losses due to calcium lines in RADYN. Here, the term ``detailed radiative losses'' describes the losses that are computed from the full NLTE treatment (i.e. lines and continua from hydrogen, helium, and calcium).
The absolute value is used inside the summations here as many of the terms inside the summation are otherwise of opposing sign and the denominator terms may be significantly smaller than the numerator, making it hard to evaluate the magnitude of these effects.

In the less energetic simulation there is a significant difference in the calcium losses between the two treatments during heating (as can be expected from the large variation in line profiles). There is also a large effect present in the F10 simulation, in a narrower band, centred on $\sim$\SI{1.2}{\mega\m} where the core of the line forms.

In both cases the region where the two treatments agree quite well ($0.6$ to \SI{1.1}{\mega\m}), is also the region where calcium losses are maximal for most of the simulation. In the $1$ to \SI{2}{\mega\m} region, the calcium lines typically represent 5-20\% of the total detailed losses, and the relative change of radiative loss between the two calcium treatments is mostly in excess of 50\%.

\begin{figure*}
    \centering
    \includegraphics[width=0.8\textwidth]{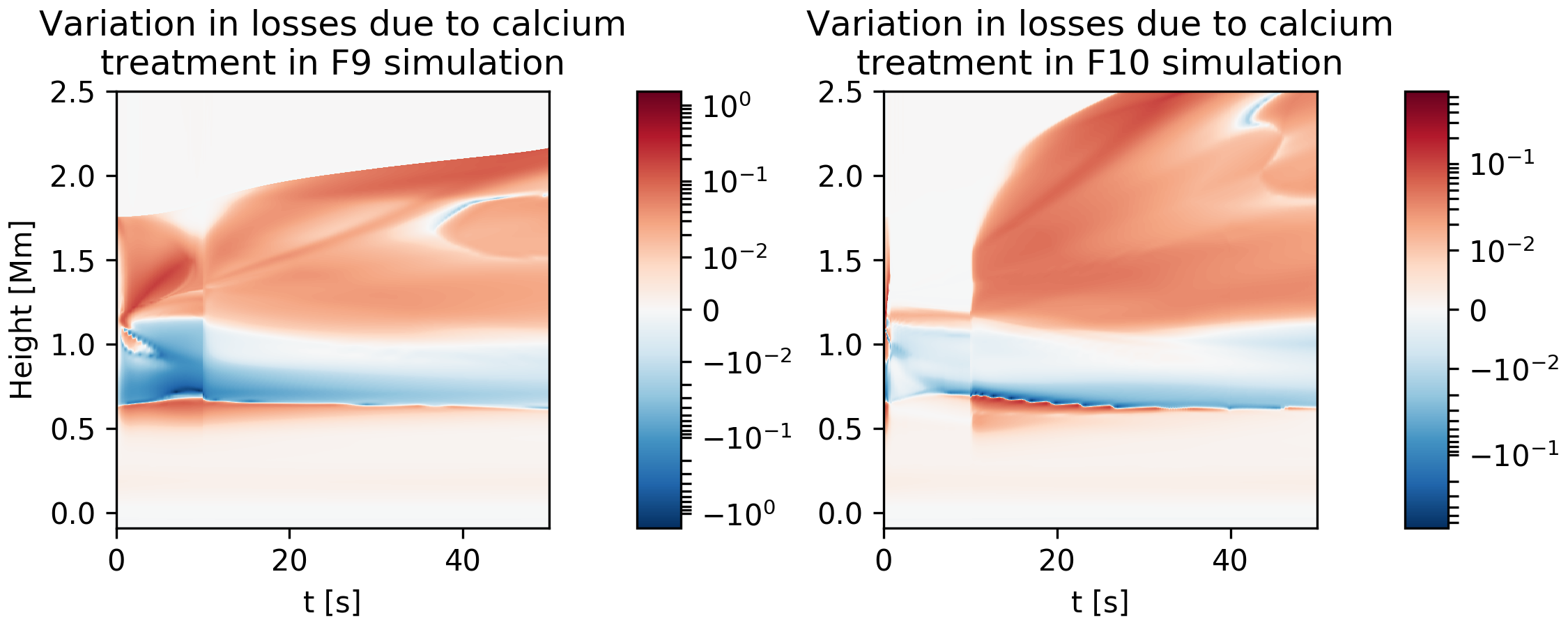}
    \caption{The absolute relative change of the detailed radiative losses, based on the change between the LE and the LI treatments (i.e. the product of the left- and right-hand panels of Figs~\ref{Fig:Losses1e9} and \ref{Fig:Losses1e10}) for both the F9 and F10 simulations.}
    \label{Fig:LossComparison}
\end{figure*}

The effect on the total detailed radiative losses in the simulation can be
approximately described by the product of the left- and right-hand panels in
Figs.~\ref{Fig:Losses1e9} and \ref{Fig:Losses1e10}. These are plotted in Fig.
\ref{Fig:LossComparison}, and there is up to a 15\% difference in the detailed
radiative losses throughout the majority of the chromosphere. The time-averaged
variation in radiative losses in the chromosphere is smaller in the F10
simulation than the F9, although they remain of the same order of magnitude, and
this is likely due to the calcium accounting for a smaller portion of the total
radiative losses in the F10 case.  It is therefore reasonable to suggest that a
complete RADYN-style simulation using an LI treatment could see changes in the
chromospheric energy balance of around 10-15\% and could therefore present
different atmospheric evolution, beyond the shape and interpretation of the
calcium line profiles.

\section{Discussion and Conclusions}

We have shown that the inclusion of the radiation field from the hydrogen Lyman lines in the calculation of the calcium populations substantially change both the emergent calcium line profiles, and the formation regions of these lines. Additionally, these effects can change the energy balance in the upper chromosphere by up to 15\% in the simulations investigated here, and could therefore modify the atmospheric evolution, further increasing the change in the emergent line profiles. It is therefore important to consider these effects and analyse them further in simulations including the energy balance from the Lyman line inclusive treatment in a self-consistent way.

Whilst not presented here, we have additionally confirmed these results using the flare simulation presented in \citet{Kasparova2019}.
The FLARIX results were inserted into a \Lw{} model and similar relative differences between LI and LE for the \CaLine{} line were found.
Both time-dependent and statistical equilibrium solutions from RADYN and FLARIX snapshots were computed for this simulation and the others presented above. Little difference was found for the \Caii{} line profiles, showing that the non-equilibrium ionisation of \Caii{} plays a marginal role here, and that of photoionisation by the hydrogen Lyman lines appears significantly greater.
There is no clear correlation between the changes to these spectral lines and the increasing flux between the F9 and F10 models, in both cases the effects are significant and appear likely to remain so as the energy input increases further.

For the simulations presented here we have also performed a cursory investigation of these effects on the \Caii{} H resonance line, and found that they are less significant than on \CaLine{}, settling back to the LE solution much faster after the beam heating ends. This may be due to the lack of overlap between Ly$\alpha$ and the \Caii{} resonance continuum, but further investigation is needed. We note that similar effects will also occur with the photoionisation of Mg\,\textsc{ii} by the Lyman continuum, but as the edge of the Mg\,\textsc{ii} resonance continuum lies at \SI{82.46}{\nano\metre}, only the Lyman continuum can contribute and the effects of the Lyman lines investigated here only apply to subordinate continua.

\subsection*{Acknowledgments}

C.M.J.O. acknowledges support from the UK Research and Innovation's Science and
Technology Facilities Council (STFC) doctoral training grant ST/R504750/1.  P.H.
and J.K. were supported by grant 19-09489S (GACR) and project RVO:67985815.  L.F.
acknowledges support from UK Research and Innovation’s Science and Technology
Facilities Council under grant award number ST/T000422/1.  This research arose
from discussions held at and around a meeting of the International Space Science
Institute (ISSI) team: ``Interrogating Field-Aligned Solar Flare Models:
Comparing, Contrasting and Improving'' organised by G.S. Kerr and V. Polito and
we would like to thank ISSI-Bern for supporting this team.